# Numerical weather prediction or stochastic modeling: an objective criterion of choice for the global radiation forecasting

Cyril Voyant[1,2], Gilles Notton[1], Christophe Paoli[3], Marie-Laure Nivet[1], Marc Muselli[1], Kahina Dahmani[4]

[1]CNRS UMR SPE 6134, University of Corsica, Ajaccio, France

[2]Hospital of Castelluccio, radiophysics unit, Ajaccio, France

[3]Galatasaray University, Computer Science Department, Çırağan Cad. No:36, Ortaköy 34357, Istanbul, Turkey

[4]Laboratory of energy mechanic and conversion systems, University HouaryBoumédiène, Algier, Algeria

Numerous methods exist and were developed for global radiation forecasting. The two most popular types are the numerical weather predictions (NWP) and the predictions usingstochastic approaches. We propose to compute a parameter noted $\chi$ constructed in part from the mutual information which is a quantity that measures the mutual dependence of two variables. Both of these are calculated with the objective to establish the more relevant method between NWP and stochastic models concerning the current problem.

## I. Introduction

Because of their random and intermittent trend, the integration rate of renewable energies is limited in the electrical grid. As a matter of fact, a limitation allows protecting the grid and to warrant a good supply

*Numerical weather prediction or stochastic modeling: an objective criterion of choice for the global radiation forecasting*

quality and to guaranty perfect production/consumption balance [Mellit, 2009; Paoli, 2010; Voyant, 2011, 2012;]. For increasing this insertion rate numerous solutions are studied and applied. Among these solutions, the main one consists in coupling the renewable energies with storage means (as hydrogen, battery, etc.). However, this coupling is not sufficient if the storage management is not well mastered. To do it, it is essential to anticipate the renewable energies production. Considering the grid manager's point of view (Köpken et al., 2004), needs in terms of prediction can be distinguished according to the considered horizon: following days, next day by hourly step, next hour and next few minutes. With efficient prediction tools dedicated to grid managers, the PV part in the mix energy would be increased; actually in France, the intermittent energy contribution is limited to 30%.Several methods exist and have been developed for twenty years: numerical weather predictions (NWP) and predictions based on stochastic approaches [Voyant, 2012; Sfetsos, 2000]. If these two methods use mathematical approaches, the first one models the atmosphere and oceans and allows predicting the weather from satellite images and primitive equations (nonlinear partial differential equations impossible to solve exactly through analytical methods) [Paulescu, 2013]. The second ones consists in statistical models allowing generating alternative versions of the time series, representing what might happen over non-specific time-periods in the future [Voyant, 2011]. The choice between these two methodologies is not really scientifically established and the purpose of this paper is to propose an objective rule allowing guiding researchers or manager in their choices related to location and spatial and temporal resolution.

This paper is organized as follows. The next section is dedicated to a brief description of the two types of global radiation prediction models, the NWP and the stochastic approaches. The section 3 introduces the methodology and the parameters proposed. Section 4 describes results and finally section 5 presents our conclusions and gives some perspectives to this work.

## II. Prediction Models

The next part describes the two available types of models for the global radiation forecasting: the NWP and stochastic approaches.

### a) NWP models

A numerical weather model is a computer program that simulates the atmospheric motion in space and time. A variety of weather phenomena can be analyzed and predicted by these types of NWP models. In this type of model, the atmosphere is represented by a 3D grid. The finer is the grid spacing the more elaborate is the simulation. The simulation by this model generates the future state of the atmosphere in each network points from its initial state [Radnoti, 1995; Voyant, 2011; Bouttier, 2010; Yessad, 2010]. Among all the NWP models, it may be mentioned weather research and forecasting model (WRF), AROME which concerns the mesoscale, or the model of the European centre for medium-range weather forecasts (ECMWF). The prediction errors of this model depend on the considered locations and fluctuate between 20 and 40% (nRMSE= normalized Root Mean Square Error, [6]).

### b) Stochastic models

The global radiation forecasting is the name given to the process used to predict the available amount of solar energy. Numerous predictive methods have been developed by experts around the world. Often the times series (TS) mathematical formalism is used [Sfetsos, 2000; Mellit, 2009; Paoli, 2010; Voyant, 2011 and 2012]. It is described by sets of numbers that measures the status of some activity over time. It is a collection of time ordered observations $x_t$, each one being recorded at a specific time t (period). A TS model ($\hat{x}_t$) assumes that past patterns will occur in the future. TS prediction or TS forecasting takes an existing series of data $x_{t-k}, .. , x_{t-2}, x_{t-1}$ and forecasts the $x_t$ data values. The goal is to

*Numerical weather prediction or stochastic modeling: an objective criterion of choice for the global radiation forecasting*

observe or to model the existing data series to enable future unknown data values to be forecasted accurately. Thus a prediction $\hat{x}_t$ can be expressed as a function of the recent history of the time series, $\hat{x}_t = f(x_{t-1}, x_{t-2}, ...x_{t-k})$. It is demonstrated that models (artificial neural networks called ANN, AutoRegressive and Moving Average called ARMA, k Nearest Neighbor called k-NN, Markov Chains, etc.) with endogenous inputs made stationary and exogenous inputs (meteorological data) can forecast the global solar radiation time series with acceptable errors [Mellit, 2009; Paoli, 2010]. Note that all the stochastic models are not equivalent, depending on the problem to solve (time horizon, spatial characteristics, locations, etc.) the best predictor to use can change. The absolute prioritization of these models is not possible.

### c) **Methodology**

Our objective is to determine the rule optimizing the choice between NWP and stochastic models considering a given spatial and temporal resolution and a location. To reach this objective we propose to compute a parameter noted $\chi$. It is constructed from the ratio between two sub-parameters, the first is related to the distances for which the global radiation series are independent ($\delta$) and the second one from a time lag for which there is no mutual dependence ($\tau$).

$$\chi = \frac{\delta}{\tau}$$

(1)

We will see in the following how to compute these two sub-parameters, but at first let us consider $\chi$. We can highlight that the dimension of this parameter $\chi$, is unconventional: pixel.time_lag$^{-1}$. The interpretation of this parameter should indicate if, for a given spatial and temporal resolution and for a given location, the TS formalism and the NWP are relevant. The most important parameter for stochastic models seems to be $\tau$: more it is important and more the information contained in the past series (intrinsic behavior) will be used to model the future. Concerning the global model NWP, it is mainly the spatial link between the points of the mesh which will guarantee the merits of the approach. In fact, the time dependence and the spatial dependence are linked, but the temporal aspect is certainly to a lesser extent. The kinetic of the cloud cover (related to the primitive equation and nonlinear partial differential equations) must be observed pixel by pixel in order to develop relevant model. Indeed, if the distance between mesh grid points is too high, local phenomena (cumulus cloud has a scale of less than 1 km) are not taken into account. But as all the pixels will be with the same average cloud cover, NWP will be very relevant in this case. For the high resolutions, the clouds local displacements appear: high heterogeneities between pixels are generated. Moreover, according to the dynamic and chaotic appearance of the fluid dynamics equations involved in weather forecasting, their modeling is very difficult and even impossible. Referring to the Equation (1), this spatial link will be illustrated by $\delta$. Thus if $\chi << 1$ stochastic models will be preferable and if $\chi >> 1$ NWP models will be more appropriate. We must note that considering the location, the time lag and the spatial resolution, $\chi$ can fluctuate significantly.

The question remains of how to find the values of $\delta$ and $\tau$. We propose to use the mutual information tool (*MI(X,Y)* in (2)) which is a quantity measuring the mutual dependence of two variables *X* and *Y*. In fact, this formalism replaces and generalizes the cross or auto-correlation and classical variogram concepts which allow to measures only the linear relationship between two variables *X* and *Y* (Pearson correlation and variogram) or two elements of time series *X* and $L^i$ *X* (*L* and *i* the lag operator and associated order ; partial or normal autocorrelation factor). Mutual information is more general and measures the reduction of uncertainty in Y after observing X. So *MI* [Kuijper, 2004] can measure non-monotonic and other more complicated relationships. It can be expressed as a combination of marginal and conditional entropies (respectively $H(X)$ and $H(X|Y)$) [Lauret, 2013].

$$MI(X,Y) = H(X) - H(X|Y) \qquad (2)$$

For the value of $\delta$ we suggest computing and analyzing the mutual information between the global radiation and the distance between considered points and for the value of $\tau$. We will focus on the mutual information between the global radiation and the time lag for the considered location. Entropy corresponds to a measure of unpredictability or information content and can be written by the following expression (entropy of a discrete random variable $X$ with possible values $\{x_1,..x_n\}$):

$$H(X) = -\sum_{x} p(x) \log(p(x)) \qquad (3)$$

One may also define the conditional entropy of two events *X* and *Y* (Equation (4)). This quantity should be understood as the amount of randomness of the random variable X given that you know the value of Y.

$$H(X|Y) = \sum_x \sum_y p(x,y) \log(p(x)/p(x,y)) \tag{4}$$

The definition of the joint probability distribution function (*p(x,y)*) and marginal probabilities (*p(x)* and *p(y)*), allows to define a new form of the mutual information as described in the equation (5).

$$MI(X,Y) = \sum_y \sum_x p(x,y) \log((p(x,y)/(p(x)p(y)))) \tag{5}$$

As the unit of the mutual information is often the bit (if the binary logarithm is used), thus it is possible to normalize this quantity by his maximal value to obtain a percentage. The new parameter is called the normalized mutual information (*nMI*, Equation (6)).

$$nMI(X,Y) = MI(X,Y)/MI(X,X) = MI(X,Y)/H(X) \tag{6}$$

## III. Results

The mathematical formalism is being enounced; the result concerning the spatial and temporal dependence of the global radiation in Corsica Island is exposed in this part. To illustrate the previous methodologies we decided to use in this study the HelioClim-3 database (HC-3). HelioClim is a family of databases containing solar irradiance and irradiation values at ground level. We have irradiance images (at an hourly step ; Wh/m²) over 8 years (2005-2012) for overall Corsica (1150 points of measurement, around 42°1'N and 9°E, with a pixel area lower than 6.5km², that is to say a grid spacing distance of 2.5 km) like seen in Figure 1 and Figure 2.

Figure 1.Points of measurements meshgrid, circle locating Ajaccio (41°55'N and 8°44'E, elev. 0-787 m), square locating Corte (42°18'N and 9°09'E, elev. 300-2626 m) and triangle Bastia (42°42'N ; 9°27'E ; elev. 0 m)

The global radiation time series used are not related to measurements but to a computing. A cross comparison between HC-3 and ground measurements in Ajaccio, Corte and Bastia is essential to validate the available data. Estimations have been evaluated in the case of Corsica, in terms of normalized Mean Bias Error (nMBE, [%]) and normalized Root Mean Square Error (nRMSE, [%]). They have beencompared to hourly pyranometrical measurementsprovided by three meteorological stations in Corsica that cover 3 years (2004 to 2006). The biases (nMBE)is negative for all the stations showing that the radiations are underestimated (between -2% and -8%).Finally yearly nRMSE are between 19.8% and 23.5 %. Despite these uncertainties, satellite estimations constitute a good alternative to ground measurements, since the meteorological public network is composed of only 6 pyranometers scattered on the territory.

Figure 1. Irradiance map computed during spring 2012 in Corsica (HC-3). The unit of the color map is Wh/m²

### a) **Spatial dependence**

In Figure 3 is represented the *nMI* versus the distance between two points. All the points of the meshing for the 8 years are used. Note that here the computing is realized for overall data, but the same type of approach can be performed season by season and area by area to improve the model and to regroup (clustering) the area with the same characteristics.

Figure 2. Normalized mutual information of the global radiation versus the distance between considered points

We see that for a distance lower than 5 km, the *nMI* value is high, but it decreases after 5 km and it remains stable (~0.85%). The space parameter ($\delta$) is determined by the intersection between the limit of *nMi* and the tangent of the fitted curve at 0 (see Fig. 2). The chosen fitted curve is an exponential decay. In fact, the limit is close to the pixel size: $\delta \sim$ 2.5km. But for other countries (and so climates), for other time steps and for other mesh grid resolutions, nothing suggests that this 1 pixel limit will be maintained. Beyond this threshold, we can consider that the global radiations received on the other pixels are independent.

### b) Temporal dependence

In Figure 4, we see for two points located in Ajaccio and in Corte, the *MI* versus the time lag (auto-mutual information). The first location is a seaside site and the second one is a mountainous site.

Figure 3. Mutual information of the hourly global irradiation versus the time lag for a 2 given points of the grid: Ajaccio (top) and Corte (bottom)

In this example, we see that the first minimum is obtained for the 9[th] and the 10[th] time lag, corresponding to 9 and 10 hours (as a complementary, for Corte the first minimum is 7 hours). Consequently using a stochastic model, for these localizations is inconsistent to predict the global irradiation at a moment t from data collected 9 and 10 hours before (in case of 1 hour time step and measured time series, see [Voyant, 2012]). If we compute the auto-mutual information for all the points of the meshing (1150 pixels), we see that the median value is 7 (min=5, max=10, mean value=7.63 and standard deviation=1.08). For the overall territory, $\tau$ is around 7 time steps i.e. 7 hours.

### c) Interpretation

In this study (overall territory) and with the methodology used, $\tau$ is 7 time lags and $\delta$ is 1 pixel and consequently $\chi = 1/7 \ll 1$. In the Ajaccio and Corte cases, the $\chi$ parameter is also less than 1 (respectively 0.1 and 0.11 pixel.time_lag$^{-1}$). As said in the beginning of this paper (section 3), we can estimate that, in considering a hourly time-step and for a spatial resolution of 2.5 km, the use of a stochastic model is relevant. In the literature, this aspect is confirmed if the stochastic model gives a nRMSE close to 15% [Sfetsos, 2000; Voyant, 2011 and 2012; paoli, 2012], NWP give less pertinent results (nRMSE> 20% [Paulescu, 2013]). If the spatial resolution increases, $\delta$ will also increase and $\chi$ will be able to be close to one or more. If now we consider a prediction of the global radiation 24 hours ahead by hourly step for example, there are always two methodologies: the NWP or the stochastic approaches. The two methods are able to propose results more or less relevant, but to make a choice between the two forecaster types without test them (it is long and laborious), it is also possible to use the ad-hoc index defined in this paper ($\chi$). We know that $\tau$ corresponds to 7 time lags so the stochastic model will not be certainly relevant for the 24 hours horizons, however it no sure that the NWP model will be more efficient. In this case $\delta$ is yet

*Numerical weather prediction or stochastic modeling: an objective criterion of choice for the global radiation forecasting*

1 pixel. Note that in the 24 hours ahead case, the prediction during the night of the global radiation does not make sense, the night values can be deleted and the 24 hours prediction can be replaced be a 12 hours predictions and the global series by a new series where only the hours between 8AM (included) and 7PM are considered. Note that this approximation induces that during summer days, the hours of sunrise and sunset are not considered. According to the Figure 4, the maximum lag to considered is yet equal to 7 hours as for the 1 hour horizon (9 and 10 hours for Ajaccio and Corte points), the data are the same thus, as the horizon of prediction is equivalent to 12 hours, the $\tau$ time lag is equivalent to 0.6 time lags (~7/12). In this configuration, the index becomes higher than 1 ($\chi=1/0.6=1.67$ pixel.time_lag$^{-1}$ for the overall territory, 1.2 pixel.time_lag$^{-1}$ for Ajaccio and 1.3 pixel.time_lag$^{-1}$ for Corte). The proposed interpretation induced that the stochastic models become not relevant, and the use of NWP is recommended. This $\chi$ interpretation is verified in [Voyant, 2013] and in [Paulescu, 2013]. In the first reference the stochastic models (MLP and ARMA) give an nRMSE close to 30% in the h+24 case and the second anRMSE between 10 and 40% (average 25%) for the NWP model and this horizon. This methodology is relevant for other horizon, other mesh grids and other locations.

## IV. Conclusion

In this paper, we proposed a methodology to justify the use of a NWP or a stochastic model according to three considered parameters: spatial resolution, temporal step and location. The mutual information and the three proposed parameters are the mathematical tools used as choice criterion between forecasting methodologies. For our case study (Corsica Island), we see that for a temporal resolution of 1 hour and a spatial resolution of 2.5km, a stochastic model is the best choice, but in the 24 hours head forecasting (prediction by hourly step for the next day) the

*Numerical weather prediction or stochastic modeling: an objective criterion of choice for the global radiation forecasting*

NWP forecasting based are the most relevant. In view to generalize this described methodology, we must validate it in other locations and for various spatial and temporal resolutions. Moreover a regional scheme will be performed, in order to separate the different microclimates and to develop model able to consider the very high resolutions of the non-hydrostatic models (as Meso-NH for example and the sub-kilometric meshgrid). In this case, a new formalism should be developed to take into account the fact that the spatial *MI* is non-continuous at the origin (nugget effect) and because the operated normalization (*nMI*) generates an offset of the curve. Another important improvement should be operated to clarify the threshold of $\chi$, the limit fixed to 1 is not clearly proven: what happens for a value of $\chi$ between 0.9 and 1.1? Is it really possible to conclude? It will be the objectives of a future paper.

*Numerical weather prediction or stochastic modeling: an objective criterion of choice for the global radiation forecasting*

## VI. Conflits of interest

None

## VII. Acknowledgement

Thanks to Armines and to the SoDa community for making the HelioClim data available (http://www.helioclim.org/index.html).

# VIII. Figures

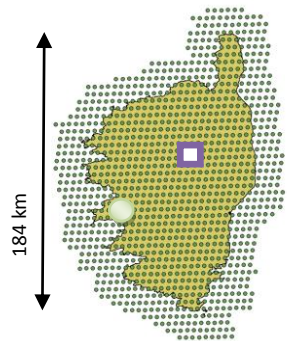

Figure 4.Points of measurements meshgrid, circle locating Ajaccio (41°55'N and 8°44'E, elev. 0-787 m), square locating Corte (42°18'N and 9°09'E, elev. 300-2626 m) and triangle Bastia (42°42'N ; 9°27'E ; elev. 0 m)

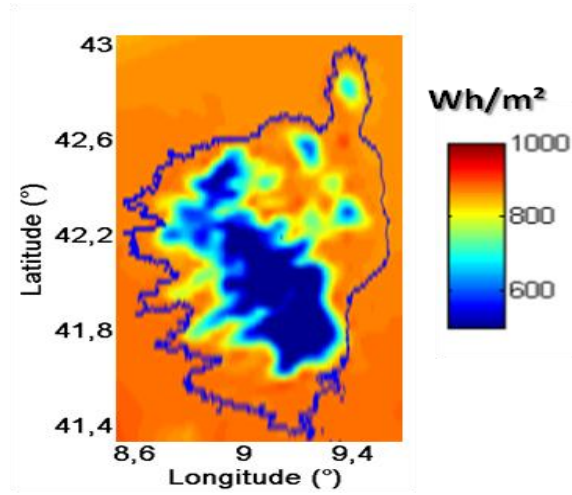

Figure 5. Irradiance map computed during spring 2012 in Corsica (HC-3).
The unit of the color map is Wh/m²

*Numerical weather prediction or stochastic modeling: an objective criterion of choice for the global radiation forecasting*

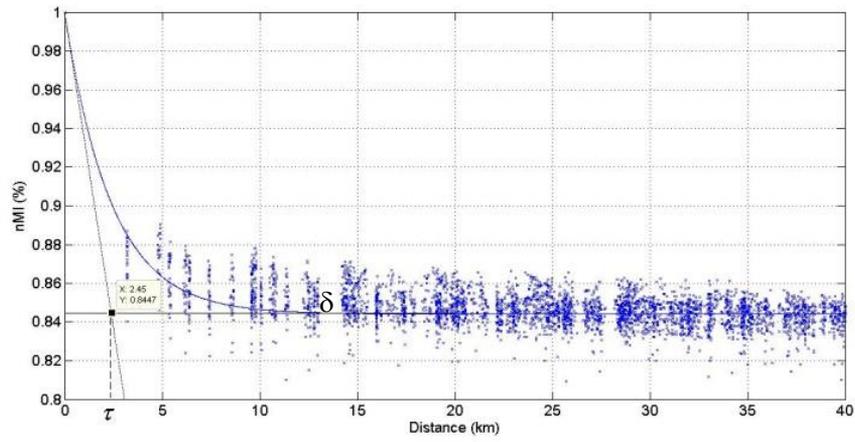

Figure 6. Normalized mutual information of the global radiation versus the distance between considered points

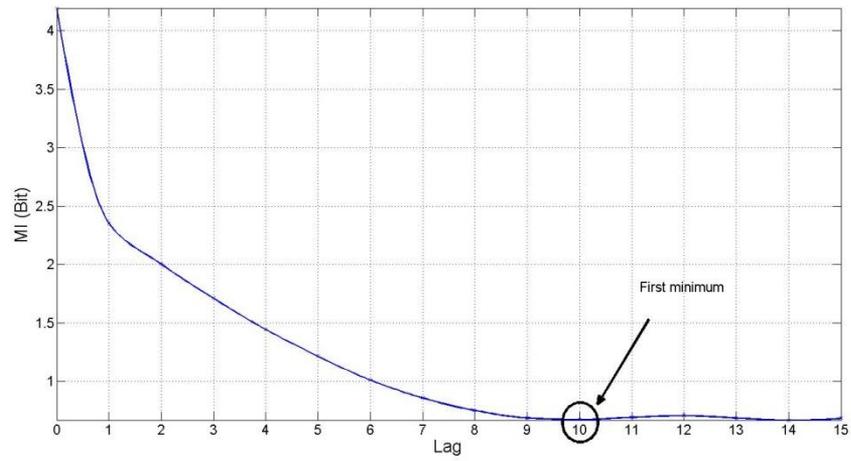

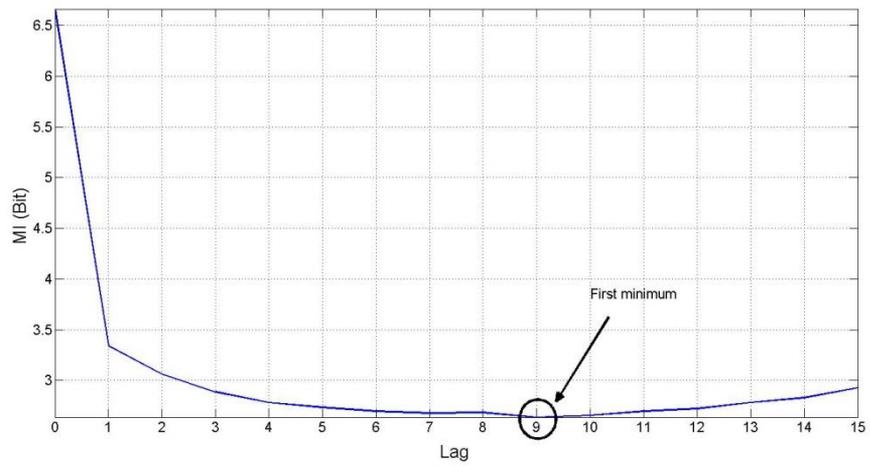

Figure 7. Mutual information of the hourly global irradiation versus the time lag for a 2 given points of the grid: Ajaccio (top) and Corte (bottom)